\documentclass[10pt, twocolumn]{article}
\usepackage{titlesec}
\usepackage{setspace}
\usepackage{enumitem}
\usepackage[font=footnotesize]{caption}
\usepackage{amssymb}
\usepackage{float}
\usepackage{amsmath}
\usepackage{indentfirst}
\usepackage{xcolor}

\titleformat*{\section}{\normalsize\normalfont\center}
\titlespacing*{\section}{1mm}{1mm}{1mm}
\titleformat*{\subsection}{\Large\bfseries}
\titleformat*{\subsubsection}{\large\bfseries}
\titleformat*{\paragraph}{\large\bfseries}
\titleformat*{\subparagraph}{\large\bfseries}

\titlespacing*{\section}
  {0pt}{1.1\baselineskip}{0.16\baselineskip}

\usepackage{scicite}

\usepackage{graphicx} 
\usepackage{times}
\usepackage{float}
\usepackage[bottom=-10cm]{geometry}

\title{DONEX: Real-time occupancy grid based dynamic echo classification for 3D point cloud}

\author{
  \normalsize{Niklas Stralau}\\\vspace{-0.2em}
  \normalsize{\textit{Department of LiDAR System Engineering}}\\ \vspace{-0.2em}
  \normalsize{\textit{Robert Bosch GmbH}}\\\vspace{-0.2em}
  \normalsize{Schwieberdingen, Germany}\\\vspace{-0.2em}
  \normalsize{Niklas.Stralau@de.bosch.com}
  \and
 \normalsize{Chengxuan Fu}\\\vspace{-0.2em}
  \normalsize{\textit{Department of LiDAR System Engineering}}\\\vspace{-0.2em}
  \normalsize{\textit{Robert Bosch GmbH}}\\\vspace{-0.2em}
  \normalsize{Schwieberdingen, Germany}\\\vspace{-0.2em}
  \normalsize{Chengxuan.Fu@de.bosch.com}
  
}

\date{}

\begin{document} 


\baselineskip24pt
\maketitle 
\begin{spacing}{1.2}

\raggedbottom


   \textbf{ ABSTRACT - For driving assistance and autonomous driving systems, it is important to differentiate between dynamic objects such as moving vehicles and static objects such as guard rails. Among all the sensor modalities, RADAR and FMCW LiDAR can provide information regarding the motion state of the raw measurement data. On the other hand, perception pipelines using measurement data from ToF LiDAR typically can only differentiate between dynamic and static states on the object level. In this work, a new algorithm called DONEX was developed to classify the motion state of 3D LiDAR point cloud echoes using an occupancy grid approach. Through algorithmic improvements, e.g. 2D grid approach, it was possible to reduce the runtime. Scenarios, in which the measuring sensor is located in a moving vehicle, were also considered.
}


\section{INTRODUCTION}
\indent To fully realize autonomous driving, the driving system must both take over all aspects of the driving function in a fully automated manner as well as serve as a fallback level. The Society of Automotive Engineers (SAE) defined five levels of autonomous driving [1]: Level 5 represents fully automated driving. Starting from SAE level-3, redundancy of sensor data would be necessary to ensure the system's functional safety. LiDAR systems act as a key technology here, complementing camera and RADAR data to ensure the safety of the system, since the driver is allowed to put his eyes off the road but must be able to take back control within seconds after a warning. LiDAR sensors can measure the environment, which is represented by 3D point clouds consisting of vast number of echoes.\\ \indent For autonomous driving functions such as braking, evasion, overtaking etc., the identification of an objects' motion states is necessary. Typically the motion state estimation of objects is done during or after the object tracking  stage of a LiDAR perception pipeline. This  kind of architecture has several drawbacks. First of all, due to occlusions and different viewing perspectives during driving, the association and motion estimation on object level or cluster level sometimes lead to false-positives, especially on static objects with irregular shapes such as vegetations. Secondly, due to the nature of the algorithms, motion estimation on object level is relatively hard to be accelerated on dedicated hardware such as GPU. Last but not least, knowing the motion state of each echo would be beneficial for different detection tasks such as free space detection, static land mark detection etc. \\ \indent In this paper, we propose a novel voxel grid based motion estimation approach for 3D LiDAR point cloud. The algorithm is called DONEX - \textbf{D}etection \textbf{O}f \textbf{N}ew \textbf{E}chos in vo\textbf{X}el grid. By comparing the volumetric occupancy of the voxels over time, dynamic echoes are detected. In addition to the method using 3D voxels, a 2D grid based method was developed to optimize the runtime. We also developed ego motion compensation so that the algorithm can cope with moving ego vehicles.\\ \indent
Our algorithm can also be used for 3D point cloud provided by other sensor modalities such as video camera or RADAR, because it only requires the cartesian coordinates of the echoes and the scan identifier. Figure 1 shows an example of the result, dynamic echoes are marked in magenta. 
\begin{figure}[H]
  \centering
      \includegraphics[width=0.5\textwidth]{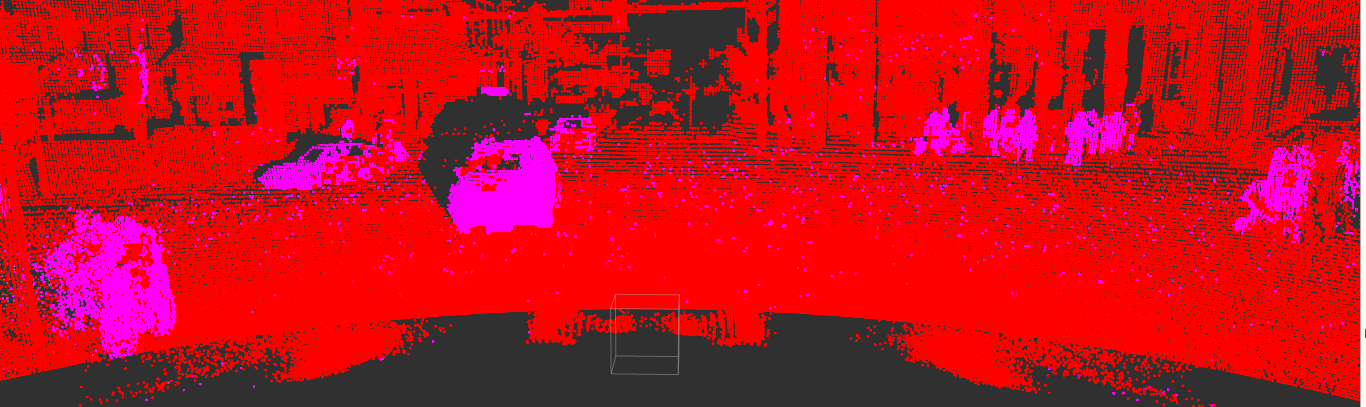}
    \footnotesize{\caption{Example of the detection results of DONEX. Identifying moving cars, pedestrians and bicycles as dynamic. }}
\end{figure}

\section{RELATED WORK}
\indent The presented solution falls into the category of 3D change detection [2]. This means time series data analysis in 3D using volumetric dynamics.
Few publications deal with the classification of dynamic and static echoes. Vieira et al [3], which use spatial density patterns, compare incoming geometries or point cloud changes soley due to occlusion or incomplete sensor coverage.\\ \indent
Another approach is the so-called fast motion segmentation by Jo et al [4]. The algorithm segments a LiDAR point cloud into dynamic and static echoes. In order to accurately and reliably estimate the motion state of each LiDAR point while considering the measurement uncertainty, both probability theory and evidence theory are used in this segmentation algorithm. The principle of probability theory is based on the geometric relationship of two
successive point clouds; the basis is a probability field, which indicates a echo as static or dynamic. If a echoes lies outside this field, an evidence-based approach - the so-called Dempster and Shafer theory [5] - is used for processing. Problems of this ray tracing approach are noises in vegetation and thin objects as well as errors at object boundaries.\\ \indent
Asvadi et al [6] also use probability theory but additionally in combination with a voxel grid. A voxel grid cell has certain probabilities to be static or dynamic, which are updated continuously. The first step is to check if any cell contains echoes. If so, it is checked whether new ones have been added in the current frame. If new echoes were detected, the dynamic probability of the voxel cell increases. If no new echoes were detected in that cell, the probability for static is correspondingly higher. One difficulty with this approach is the shadow handling. In contrast, our DONEX algorithm does not determine the probablistic motion state of a cell, but purely through real detections. Errors that are caused by a probability model do not occur in our solution, shadow handling issues could be solved by DONEX as well. \\ \indent
Publications [7][8] also use an approach with a voxel data structure
to differentiate between static and dynamic sections of the
point cloud. Instead of measuring free voxels, the system counts how many times a voxel is occupied. Because of the variating
occlusion, they must make some environment assumptions
and apply several heuristics that are not necessary in DONEX. In addition, their approach needs an ground surface estimation.
\\ \indent
The algorithm which serves as starting point of our development is the so-called peopleremover algorithm [9]. For an existing set of 3D point clouds, a binary voxel occupancy grid is generated. This is traversed along the lines of sight between the sensor and the measured echoes to find differences in volumetric occupancy between scans. The traversal of the line of sight is based on the fast voxel traversal algorithm of Amanatides and Woo [10]. This method is used to investigate whether an echo that was at a certain location at another time is now no longer present there. The principle of see-through voxels is used for this purpose: If a line of sight can be traversed through a voxel where it could not be traversed at an earlier time due to an object, this object must have moved away and the voxel is now see-through. That is how a dynamic voxel is determined. A significant disadvantage of this algorithm is that it cannot be used in real-time. The peopleremover algorithm can only be used for post-processing of already stored point cloud data. A use in the field of automated driving is excluded.\\ \indent
One of the latest deep-learning based approaches from Sun [11] uses an extended SalsaNext semantic segmentation network, where a spatial and channel attention module is used to extract motion information from residual image, which is computed by the normalized range distances of the current frame and the projected previous frame. Despite good performance, the required runtime is however relatively large. \\ \indent
Another recent related work is a deep learning-based approach from Chen et al [12]. It exploits sequential range images from a rotating sensor as an intermediate representation and combines it with a convolutional neural network (CNN). Based on this intermediate representation, the existing range-image-based semantic segmentation networks can directly be exploited as already proposed for example by Milioto et al. [13], Cortinhal et al. [14], and Li et al. [15] to deal with the moving objects segmentation. The CNN based segmentation approach uses the generated 3D range image from the sensor together with the residual images generated from past scans as inputs and outputs. The network exploits the temporal information and can differentiate between moving and static objects combining the range images and the residual images.
\section{GENERAL DESIGN}
\indent The developed algorithm DONEX uses a binary occupacy grid like the proposed peopleremover approach. The concept of occupancy grids for map generation with mobile robots was first introduced by Hans Moravec and Alberto Elfes in their oft-cited publications [16]. Figure 2 shows an example of an occupancy grid at two time stamps. The information about the measured echoes is durable stored over time. For example, we know that in t\textsubscript{n+1} in voxel (0/2) the echoes with both identifiers were already registered.
\begin{figure}[H]
  \centering
      \includegraphics[width=0.5\textwidth]{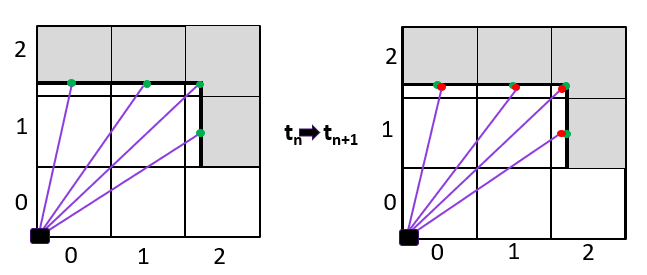}
    \footnotesize{\caption{Binary occupacy voxel grid. Gray fields represent a static object; green points are echoes of the first scan, red ones from the second scan. }}
\end{figure}  Compared to processing with raw 3D point cloud data, using an occupancy grid is significantly more runtime efficient. Neither the coordinates nor the number of echoes associated with a voxel are stored; the size of the used data is significantly smaller than that of the input data. A significant factor is the size of the voxels - smaller voxels provide a more accurate result, but require a longer runtime. In addition, the number of voxels in the grid affects the runtime. This depends on the space considered as well as on the dimension of the voxels, whereby the second point has a bigger impact.

Peopleremover identifies a voxel as dynamic if it does not detect any echoes in the current scan but it was occupied in the previous scan. Our solution follows the opposite principle: a voxel is classified as dynamic if echoes are measured but none were detected in the previous scan. In contrast to the peopleremover algorithm, this approach enables a real-time classification of the motion state of the echoes. The peopleremover can only detect that there was a dynamic echo at a certain position at an earlier time stamp.
\section{3D GRID APPROACH}
The 3D voxel method uses a grid where each echo is assigned to a voxel according to its x, y and z-coordinates. As soon as a voxel is associated with newly detected echoes (i.e. no echo was measured here in the previous scan), it is marked as dynamic. Figure 3 demonstrates this in a 2D view. 
\begin{figure}[H]
  \centering
      \includegraphics[width=0.5\textwidth]{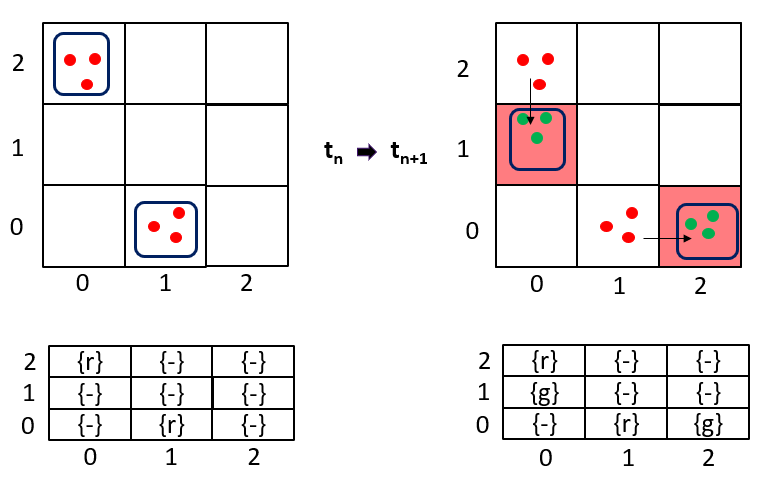}
    \footnotesize{\caption{New detection principle for 3D grid. Rectangles represent potential moving objects; movement is indicated with arrows. The grids below show the stored identiers per each voxel. Voxels identified as dynamic according to the new detection are marked in magenta. }}
\end{figure} 
In t\textsubscript{n+1} the voxels (0/1) and (2/0) have been determined to be dynamic. This can be concluded from the fact that they do not have a green identifier stored in them, since no echoes were measured in t\textsubscript{n}. A problem with this naive approach is that static objects appearing out of previously shadowed area like in the example of Figure 4. As soon as any voxel is covered by a shadow, no echo is detected. Once a voxel is no longer covered by a shadow, a new detection is always generated, which leads to false-positives for static objects. 
\begin{figure}[H]
  \centering
      \includegraphics[width=0.5\textwidth]{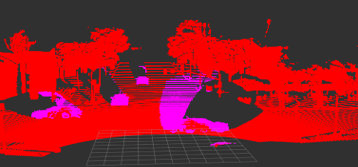}
    \footnotesize{\caption{Problem of the 3D new detection approach due to shadows. Some static objects are incorrectly classified as dynamic.}}
\end{figure} 
One approach for shadow detection from the literature is the paper by Hu and Li [17].
This is based on the concept of z-buffering known from computer graphics for the occlusion computation. Since the z-buffer algorithm is quite computationally intensive, a different approach was created for our development, the so-called concept of shadowing. It is based on the shadow cast of the objects. 
\begin{figure}[H]
  \centering
      \includegraphics[width=0.5\textwidth]{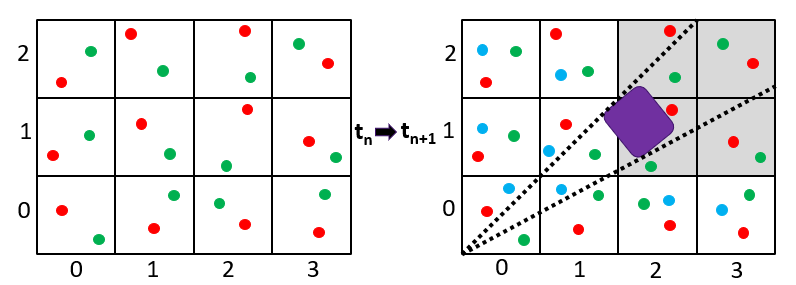}
    \footnotesize{\caption{Concept of shadowing voxels. After the scan of t\textsubscript{n}, green and red identifiers are stored in the voxels. In t\textsubscript{n+1} echoes with blue identifiers are detected. Voxels where no echo detection was measured, but one was measured in the scan before, are classified as shadowed. Corresponding voxels are marked in grey. }}
\end{figure}
However, the shadowing leads to a new problem. Once a dynamic object has traversed a voxel, the corresponding identifier is stored. After the object moves away, the voxel would be falsely shadowed. To compensate for these false-negatives, the fast voxel traversal algorithm of Amanatides and Woo [10] is used. This is performed in each scan to identify the see-through voxels as in the peopleremover algorithm [9]. These dynamic voxels are exactly the ones that can lead to potential false-negatives. To prevent this, they become ''de-shadowed''. \\ \indent Thus, false-positives and false-negatives can be prevented by a combination of shadowing and the fast-voxel-traversal algorithm. 
\section{2D GRID APPROACH}
To improve the runtime of the algorithm, another approach - based on a 2D grid - was developed. In this case, echoes are sorted into 2D cells based on their x- and y-coordindates. To compensate for the missing z-coordinate, the principle of ranging was invented. In addition to the scan identifiers, a 2D voxel stores a range consisting of two floating point numbers: The highest and the lowest z-value of all detected echoes. Thus, the range of the z-coordinates of all associated echoes are represented. An increase of the range implies a new detection or the occurrence of dynamic echoes within the 2D voxel. Figure 6 illustrates the principle. \\
\begin{figure*}[h]
  \centering
      \includegraphics[width=1\textwidth]{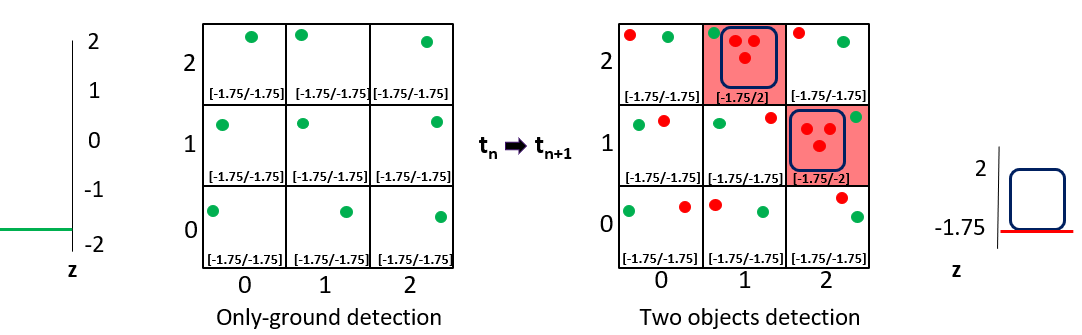}
    \footnotesize{\caption{Concept of ranging. Each voxel stores highest and lowest z-value of the detected echoes from a scan as a range. Extension of the range of a voxel implies a new detection respectively a voxel containing dynamic echoes.}}
\end{figure*}
At the time t\textsubscript{n} only the ground is measured in the whole grid. Since this is at the z-value -1.75, all 2D voxels have the range [-1.75/-1.75]. In t\textsubscript{n+1} there are now objects in (2/1) and (1/2) that have been added. These objects have a height over ground of 2m. Accordingly, the ranges of the voxels are updated to be [-1.75/2.0]. As a result of the z-range expansion, these voxels are classified as dynamic.\\ \indent
A naive 2D approach produces both, false-positives and false-negatives. False-positives are produced by labeling all echoes of a 2D voxel as dynamic. Therefore, it is necessary to recognize only those that were newly added as dynamic. A typical example of these false-positives are ground echoes. The solution to this problem is to save the previous ranges. In case of a range extension, the echoes which belong to both the previous saved range and the new range are classified as static. Newly added dynamic echoes can be determined. \\  \indent False-negatives occur when an object moves away from a 2D voxel. Since the voxel stores the z-range, smaller objects or objects of the same size that later move into this 2D voxel are incorrectly classified as static. To compensate for this, the technique of resizing was introduced: Once an object has moved away from a 2D voxel, the z-range of that 2D voxel is reset.\\ 
\section{EGO MOTION COMPENSATION}
In order to make the developed algorithm applicable to real world driving situations, the ego vehicle's motion must be taken into account. For this work, we used odometry data provided by vehicle inertia sensor which provides the linear velocity and the angular velocity - each for each of the three dimensions. As Figure 7 shows, the motion of the vehicle is therefore described by six degrees of freedom (6DOF) [18]. \\
\begin{figure}[H]
  \centering
      \includegraphics[height=0.3\textwidth]{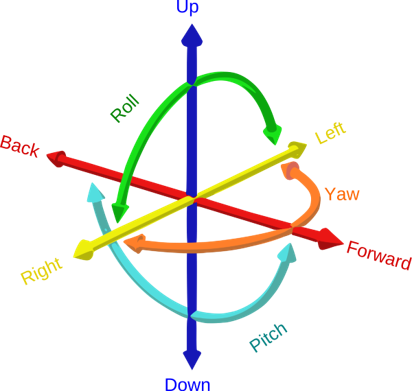}
    \footnotesize{\caption{Visualisation of the 6DOF. Consists of three angular and three linear components, one per each dimension.}}
\end{figure}
Using the odometry data, the motion of the vehicle can be compensated by a coordinate transformation of two consecutive point clouds. To convert the required velocities into absolute distances, the time difference between two frames, which is calculated by the corresponding time stamps, is used. \\ \indent
Mathematically, the transformation is done by multiplying all points with a transformation matrix. This matrix consists of the rotation matrix and the translation matrix. In order to merge these two into a homogeneous matrix, a fourth dimension must be added; the main diagonal is supplemented by a one and the remaining lower rows are filled with zeros. Equation 1 shows this.

\begin{equation}
 \begin{pmatrix}
p_{x'} \\
p_{y'} \\
p_{z'}\\
1\\
\end{pmatrix}
=
\begin{pmatrix}
r_{11} & r_{12} &  r_{13} & t_x \\
r_{21} & r_{22} &  r_{23} & t_y  \\
r_{31} & r_{32} &  r_{33} & t_z\\
0 &0 &0 &1 
\end{pmatrix}
* 
\begin{pmatrix}
p_{x} \\
p_{y} \\
p_{z}\\
1\\
\end{pmatrix}
\end{equation}
\\
\indent To integrate the ego-motion compensation into the existing grid structure, the method of voxel shifting was developed. The coordinate system and its reference is always the current frame respectively the current vehicle position. The data of the previous frame is transformed to the current frame. This is achieved by moving the voxels and their contents according to the transformation matrix. Thus, not the point cloud, but the voxels are transformed. The number of transformations is significantly lower than with a point cloud based approach, the runtime is accordingly lower. Figure 8 illustrates this concept with a simple example in two dimensions.
\begin{figure}[H]
  \centering
      \includegraphics[width=0.5\textwidth]{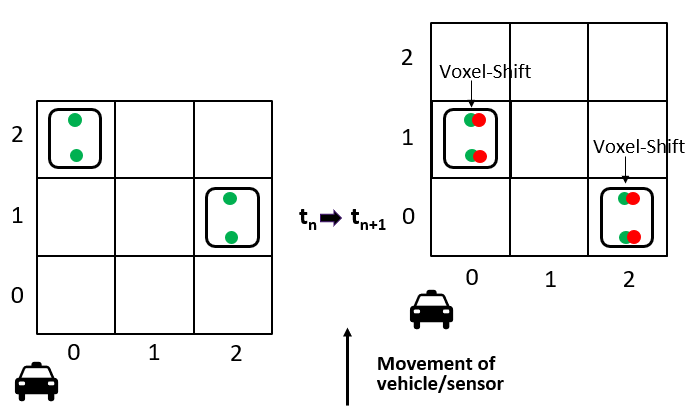}
    \footnotesize{\caption{Concept of voxel-shift. According to the transformation matrix all voxels respectively their content is shifted. The rectangle objects could correctly be classiefied as static in this example.}}
\end{figure} 
At the time t\textsubscript{n} echoes of static objects are detected in the voxels (0/2) and (2/1). The position of the echoes is relative to the sensor, which is modeled through the passenger car here. At time t\textsubscript{n+1} this car moves one voxel length in x-direction. This movement can be determined by current odometry data and a corresponding
transformation matrix can be generated. Consequently, it is possible to calculate at which position in t\textsubscript{n+1} which voxels of the reference frame t\textsubscript{n} are located. Accordingly, the voxels are moved, respectively their contents are copied. \\ \indent
The explanation in Figure 8 is a very simplified example. The movement of the vehicle is based on only one degree of freedom, because the vehicle moves only forward (x-direction). Here, a one-dimensional translation would be sufficient for compensation. Since 6DoF occur in the systems in practice, the use of a transformation matrix is necessary here.

\section{RESULTS} 
The development was done in the programming language C++. The complilation of the source code was performed in realease mode with O3 flag compiler. The algorithm has been executed on an Intel Core i7-8650U processor with 1.90 GHz on Ubuntu 18.04.4.\\ \indent
On average, there were 4-7 moving objects with a speed of about 30-60 km/h on the observed frames. The point cloud of a frame consists of about 160,000 points respectively echoes.\\ \indent \color{black}
Overall, DONEX experimental works for static and dynamic scenarios with moving ego vehicle. The algorithm was validated using different scenarios. Good results were obtained with both the 2D and 3D grid approaches.\\ \indent
One reason for developing the 2D approach was to optimize the runtime.  To determine the improvement in performance, runtime measurements were performed. For both approaches, the runtime was measured for the same voxel side lengths and the same effective grid size. Table 1 shows the results. 
\begin{table}[H]
\hspace{0.85cm}
\begin{tabular}{|ccc|cc}
\cline{1-3}
Side length [m] &Variant   & Runtime [ms] &  &  \\ \cline{1-3}
0.3 &3D              &  175 $\approx$ 185                     &  &  \\
                   0.3         &2D                        &  8 $\approx$ 9                      &  &  \\ \cline{1-3}
                    0.15 & 3D               &  550 $\approx$ 560          &  &  \\ 
                     0.15        &2D                        & 12 $\approx$ 14          &  &  \\ \cline{1-3}
                   0.1   &3D                &  1300 $\approx$ 1350      &  &  \\ 
                    0.1          &2D                        &  21 $\approx$ 23                &  &  \\ 
                    \cline{1-3}
\end{tabular}
\caption{Runtime comparison 2D grid vs. 3D grid. In these experiments we have voxelized the point clouds in the following range according x- and y-direction: xMin = 0m, xMax = 75m, yMin = -25m and yMax = 25m.}
\label{tab:my-table}
\end{table}
For the largest voxels, the difference in runtime between the 2D and 3D approaches is obvious. With decreasing voxel sizes, the runtimes increase exponentially. Especially using 3D grid, the runtimes quickly become much too high to use the algorithm in practice. The 2D approach, on the other hand, shows good performance even for small voxel sizes. Several measurements showed that the result with a voxel side length of 0.15m has the best compromise between quality of the result and runtime. The improvements of detection performance for smaller voxels are marginal and hardly visually noticeable. Figure 9 shows the results with different lengths.
\begin{figure}[H]
  \centering
      \includegraphics[width=0.5\textwidth]{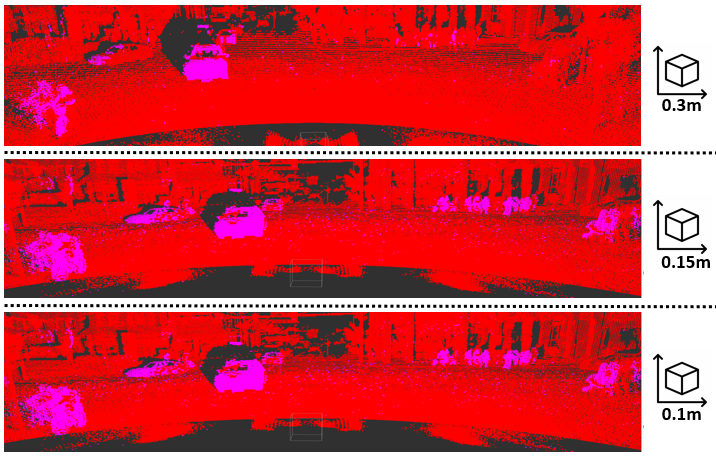}
    \footnotesize{\caption{Examples for performing DONEX algorithm with different voxel side lengths. Here we show the results using 2D grid, as the results using 3D grid are visually similar }}
\end{figure} 
Subjectively, the detection performance of the 2D and 3D approaches are quite similiar despite the huge differences in runtime. Following graphic shows an example scenario with a static ego vehicle.
 \begin{figure}[H]
  \centering
      \includegraphics[width=0.5\textwidth]{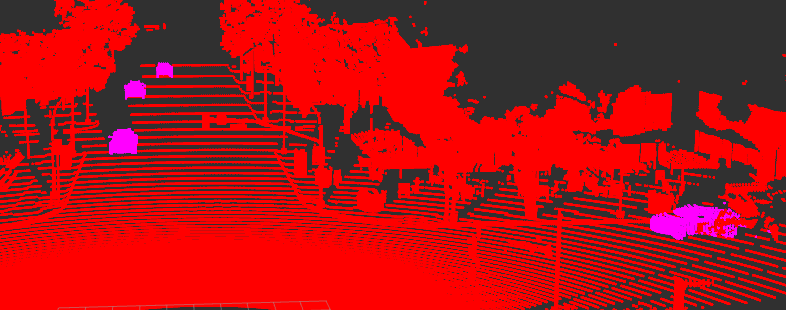}
    \footnotesize{\caption{Example of the DONEX algorithm in a static ego vehicle scenario. Identifying moving cars as dynamic}}
\end{figure}
Figure 11 demonstrates an example scenario with a moving ego vehicle. 
\begin{figure}[H]
  \centering
      \includegraphics[width=0.5\textwidth]{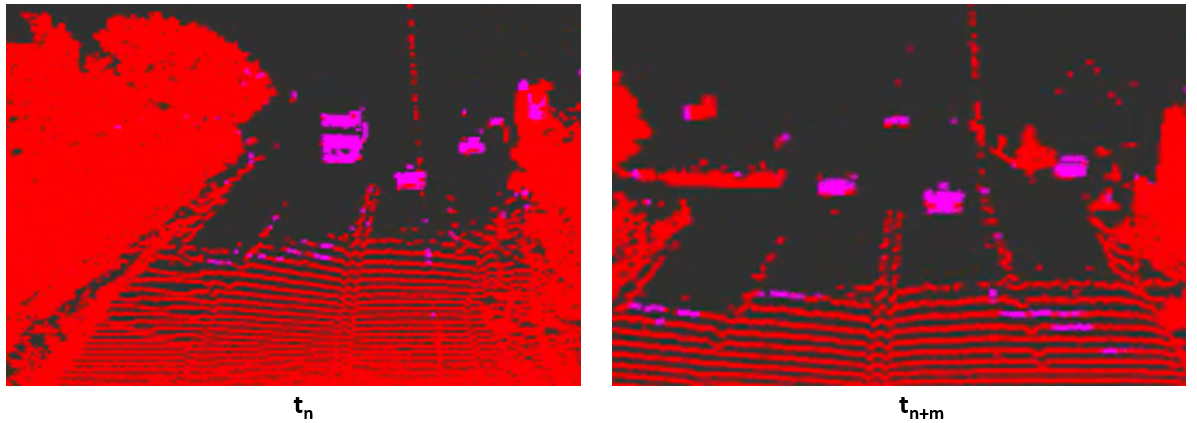}
    \footnotesize{\caption{Example for moving ego vehicle scenarios. In this example, m is 43. As the frame rate is 10Hz, the time difference between the two frames is 4.3s.}}
\end{figure} 
The newly developed DONEX algorithm is inspired by some concepts of the papers presented in related work and includes various own ideas and approaches. Table 2 shows a comparative view between DONEX, the peopleremover [9], the fast motion segmentation with ray tracing approach [4] and the motion voxel grid approach which is based on the probability theory. \\ \indent
Like peopleremover and motion voxel grid, DONEX belongs to the voxel-based approaches in contrast to ray tracing. Here, DONEX is more similar to peopleremover, since the detection is based on real detections and not on probabilities as in motion voxel grid. In contrast to the peopleremover, DONEX can also be used in real-time and has advantages in terms of runtime - in particular, the 2D implementation of DONEX offers a lot of potential to have an adaptable implementation and massively improve the runtime. \\ \indent
All voxel-based approaches have the advantage of better adaptation in contrast to the ray tracing approach, since the grid and the size of the individual voxels can be optimized according to the use case. Grid and voxel size obviously behave antiproportionally to the runtime; anyway, the development of the 2D approach brings exorbitant improvements in runtime (see Table 1).  \color{black}

\begin{table*}[t]
\hspace{0.85cm}
\begin{tabular}{l|l|l|l|l|}
\cline{2-5}
                                                                                       & DONEX                                                                                                                            & Peopleremover {[}9{]}                                                                                                             & Ray tracing {[}4{]}                                                                                          & Motion voxel grid {[}6{]}                                                                       \\ \hline
\multicolumn{1}{|l|}{\begin{tabular}[c]{@{}l@{}}Main\\ idea\end{tabular}}              & \begin{tabular}[c]{@{}l@{}}Determine if a new\\ point is detected in\\ a voxel grid cell\end{tabular}                            & \begin{tabular}[c]{@{}l@{}}Determine if a point\\ that was present at \\ an earlier view point \\ is not present now\end{tabular} & \begin{tabular}[c]{@{}l@{}}Determine how often\\ a point was observed \\ at earlier view points\end{tabular} & \begin{tabular}[c]{@{}l@{}}Update static/dynamic\\ likelihood of voxel\\ grid cell\end{tabular} \\ \hline
\multicolumn{1}{|l|}{\begin{tabular}[c]{@{}l@{}}Data\\ structure\end{tabular}}         & 2D/3D voxel grid                                                                                                                 & 3D voxel grid                                                                                                                     & Mesh grid buffer                                                                                             & 3D voxel grid                                                                                   \\ \hline
\multicolumn{1}{|l|}{\begin{tabular}[c]{@{}l@{}}Detection\\ type\end{tabular}}         & \begin{tabular}[c]{@{}l@{}}Based on real\\ detections\end{tabular}                                                               & \begin{tabular}[c]{@{}l@{}}Based on real\\ detections\end{tabular}                                                                & \begin{tabular}[c]{@{}l@{}}Based on \\ likelihood\end{tabular}                                               & \begin{tabular}[c]{@{}l@{}}Based on \\ likelihood\end{tabular}                                  \\ \hline
\multicolumn{1}{|l|}{\begin{tabular}[c]{@{}l@{}}Algorithmic \\ runtime\end{tabular}}   & Very good                                                                                                                        & Bad                                                                                                                               & Good                                                                                                         & Very good                                                                                       \\ \hline
\multicolumn{1}{|l|}{\begin{tabular}[c]{@{}l@{}}Real-time\\ usage\end{tabular}}        & Yes                                                                                                                              & \begin{tabular}[c]{@{}l@{}}No, only\\ post-processing\end{tabular}                                                                & Yes                                                                                                          & Yes                                                                                             \\ \hline
\multicolumn{1}{|l|}{\begin{tabular}[c]{@{}l@{}}Adaption\\ possibilities\end{tabular}} & \begin{tabular}[c]{@{}l@{}}Grid- and voxel size\\ is adaptable, \\ high performance\\ potential with \\ 2D approach\end{tabular} & \begin{tabular}[c]{@{}l@{}}Grid- and voxel size\\ is adaptable\end{tabular}                                                       & \begin{tabular}[c]{@{}l@{}}No adaption\\ possible\end{tabular}                                               & \begin{tabular}[c]{@{}l@{}}Grid- and voxel size\\ is adaptable\end{tabular}                     \\ \hline
\end{tabular}
\caption{Comparison between DONEX and three approaches from related papers. Quality criteria are runtime, real-time usage and adaptability }
\label{tab:my-table}
\end{table*}
\newpage
\section{FUTURE WORK}
In the future, it is planned to evaluate, design and implement further optimizations regarding the runtime. For example, using hardware based acceleration techniques, DONEX algorithm can be parallelized. Aspects such as the voxel structure as well as the management of the corresponding voxel data will be considered. Whether the use of hashmaps, for example, can further improve the runtime. \\ \indent
Furthermore, DONEX algorithm will be quantitatively more deeper compared with other solutions, both regarding runtime and the detection performance.
\newpage
\section{CONCLUSION}
In this paper, the DONEX algorithm was presented, which can effectively classify dynamic echoes in a LiDAR point cloud using an occupancy grid. Only the scan identifier and the cartesian coordinates of the individual echoes of the point cloud are required as input data. Therefore, the algorithm can be used to process 3D point cloud data provided by any other sensor modalities. \\ \indent
The 3D approach which was inspired by the peopleremover algorithm was modified by the idea of the new detection so that it can be applied in real-time systems, e.g. on an autonomous driving vehicle. A 2D variant was developed for runtime optimization. \\ \indent
We have shown that by using ego-motion compensation techniques, this algorithm can cope with moving ego vehicle.  
\end{spacing}
\clearpage
\section*{REFERENCES}
\small
\singlespace
\vspace{-0.1mm}
\begin{enumerate}[label={[\arabic*]}]
\item SAE. ''Taxonomy and Definitions for Terms Related to Driving Automation
Systems for On-Road Motor Vehicles''. In: Standard J3016 (2014).
\item R. Qin, J. Tian, P. Reinartz. “3d change detection–approaches and applications”. In: ISPRS Journal of Photogrammetry and Remote Sensing, vol. 122, pp. 41–56 (2016).
\item A. W. Vieira, P. L. Drews, M. F. Campos. “Spatial density patterns
for efficient change detection in 3d environment for autonomous
surveillance robots". In: IEEE Transactions on Automation Science and
Engineering, vol. 11, no. 3, pp. 766–774 (2014).

\item S. Lee, C. Kim, K. Jo. "Rapid Motion Segmentation of LiDAR Point Cloud
Based on a Combination of Probabilistic and Evidential Approaches for Intelligent Vehicles“. In: Multidisciplinary Digital Publishing Institute (MDPI)(2019).
\item G. Shafer. "A Mathematical Theory of Evidence“. In: Princeton University
Press (1976).
  \item A. Asvadi, C. Premebida, P. Peixoto, U. Nunes. "3D Lidar-based Static and Moving Obstacle Detection in Driving Environments:
an approach based on voxels and multi-region ground planes". In: Institute of Systems and Robotics, Department of Electrical and Computer Engineering, University of Coimbra - Polo II, 3030-290 Coimbra, Portugal (2016).
\item A. Asvadi, P. Peixoto, U. Nunes. “Two-stage static/dynamic environment modeling using voxel representation". In: Robot 2015: Second
Iberian Robotics Conference. Springer, pp. 465–476 (2016).
\item A. Asvadi, C. Premebida, P. Peixoto, U. Nunes. “3d lidar-based
static and moving obstacle detection in driving environments: An
approach based on voxels and multi-region ground planes”. In: Robotics
and Autonomous Systems, vol. 83, pp. 299–311 (2016).
  \item J. Schauer, A. Nüchter. ``The peopleremover – removing dynamic objects from 3D point cloud data by traversing a voxel occupancy grid''. In: IEEE Robotic and Automation Letters (2018).
  \newpage
  \item J. Amanatides, A. Woo et al. ''A fast voxel traversal algorithm for ray tracing''. In Eurographics, vol. 87, no. 3, pp. 3–10 (1987).
\item J. Sun, Y. Dai, X. Zhang: "Efficient Spatial-Temporal Information Fusion for LiDAR-Based 3D Moving Object Segmentation". In: arXiv preprint arXiv:2207.02201 (2022).
\item X. Chen, S. Li, B. Mersch, L. Wiesmann.                                                                                                  "Moving Object Segmentation in 3D LiDAR Data: A Learning-based Approach Exploiting Sequential Data". In: IEEE ROBOTICS AND AUTOMATION LETTERS (2022).
\item A. Milioto, I. Vizzo, J. Behley, C. Stachniss. "RangeNet++: Fast and Accurate LiDAR Semantic Segmentation". In: Proc. of the IEEE/RSJ Intl. Conf. on Intelligent Robots and Systems (2019).
\item T. Cortinhal, G. Tzelepis, E.E. Aksoy. "SalsaNext: Fast, Uncertainty-Aware Semantic Segmentation of LiDAR Point Clouds". In: Proc. of the IEEE Vehicles Symposium (IV) (2020).
\item S. Li, X. Chen, Y. Liu, D. Dai, C. Stachniss, J. Gall. "Multi-scale interaction for real-time lidar data segmentation on an embedded platform". In: arXiv preprint arXiv:2008.09162 (2020). \color{black}
\item H. Moravec, A. Elfes. "High resolution maps from wide angle sonar“. In:
Proceedings. 1985 IEEE international conference on robotics and automation,
volume 2, S.116-121 (1985).
\item X. Hu, X. Li. "Fast occlusion and shadow detection for high resolution remote sensing image combined with lidar point cloud“. In: International Archives of the Photogrammetry,
Remote Sensing and Spatial Information Sciences, Volume XXXIX-B7, 2012
XXII ISPRS Congress, 25 August – 01 September 2012, Melbourne, Australia
(2012).
\item J. Craig. "Introduction to Robotics: Mechanics and Control". In:
Addison-Wesley (1986). 

\end{enumerate}

\end{document}